\documentclass[a4paper,12pt]{article}
\usepackage{amsmath,amssymb,bm}
\usepackage[utf8x]{inputenc}
\usepackage{textcomp,marvosym}

\usepackage{graphicx}
\usepackage{tikz,pgfplots}
\usetikzlibrary{shapes,spy,calc,positioning}
\usepgfplotslibrary{groupplots}
\usepgfplotslibrary{external}
\usepackage{changepage}

\usepackage{array}

\newcolumntype{+}{!{\vrule width 2pt}}

\newlength\savedwidth

\newcommand\thickhline{\noalign{\global\savedwidth\arrayrulewidth\global\arrayrulewidth 2pt}%
\hline
\noalign{\global\arrayrulewidth\savedwidth}}

\newcommand{\includetikz}[1]{%
  \includegraphics{#1.pdf}
}

\newcommand{\citep}{\cite}

\definecolor{col1}{rgb}{0,0.4470,0.7410}
\definecolor{col2}{rgb}{0.8500,0.3250,0.0980}
\definecolor{col3}{rgb}{0.9290,0.6940,0.1250}
\definecolor{col4}{rgb}{0.4940,0.1840,0.5560}
\definecolor{col5}{rgb}{0.4660,0.6740,0.1880}
\definecolor{col6}{rgb}{0.3010,0.7450,0.9330}
\definecolor{col7}{rgb}{0.6350,0.0780,0.1840}

\title{Robust modulation of \\integrate-and-fire models}
\date{}
\author{Tomas Van Pottelbergh \and Guillaume Drion \and Rodolphe Sepulchre}

\begin{document}

\maketitle

\section{Abstract}
\label{sec:abstract}

By controlling the state of neuronal populations, neuromodulators ultimately affect behaviour. A key neuromodulation mechanism is the alteration of neuronal excitability via the modulation of ion channel expression. This type of neuromodulation is normally studied via conductance-based models, but those models are computationally challenging for large-scale network simulations needed in population studies. This paper studies the modulation properties of the Multi-Quadratic Integrate-and-Fire (MQIF) model, a generalisation of the classical Quadratic Integrate-and-Fire (QIF) model. The model is shown to combine the computational economy of integrate-and-fire modelling and the physiological interpretability of conductance-based modelling. It is therefore a good candidate for affordable computational studies of neuromodulation in large networks.



\section{Introduction}
\label{sec:introduction}

Integrate-and-fire modelling has existed since the early days of mathematical physiology~\citep{Lapicque1907} and is still popular after sixty years of physiological conductance-based modelling. It captures the hybrid nature of nerve excitability by combining a continuous-time differential equation with a discrete reset rule. The differential equation integrates the flow of ionic currents according to the continuous-time laws of electrical circuits. The reset rule accounts for the all-or-none nature of the spike. Integrate-and-fire models are cheap to simulate, allowing for network computational studies of thousands or millions of interconnected neurons, but they lack physiological interpretability. In contrast, conductance-based modelling provides a computational framework to simulate  with great biophysical realism the continuous-time flow of ionic currents. Over time, they have provided a detailed understanding of neuronal excitability, shedding light on the biophysical mechanisms that shape the action potential of a particular neuron and that continuously modulate the electrical activity of a neuron across a variety of firing patters. Conductance-based models often lead to high-dimensional nonlinear differential equations that are expensive to simulate and require the tuning of many parameters, prohibiting their use in large network computational studies.

Much research has been devoted to combine the economy of integrate-and-fire models with the physiological interpretability of conductance-based models. An early example is the Quadratic Integrate-and-Fire (QIF) model that has the interpretation of a mathematical reduction of the seminal conductance-based model of Hodgkin and Huxley~\citep{Hodgkin1952}, see for instance~\citep{Izhikevich2007}. Several generalisations of the QIF model have been studied in the literature. They include the Izhikevich model~\citep{Izhikevich2003}, the AdEx model~\citep{Brette2005} and Generalised Linear Integrate-and-Fire (GLIF) models~\citep{Mihalas2008,Mensi2012,Pozzorini2013,Mensi2016}. These models can simulate an impressive variety of firing patterns with very few tuning parameters and the computational economy of the QIF model. 

The present paper is a continuation of this work. Our aim is to make integrate-and-fire modelling suitable for neuromodulation studies, specifically to study the effect of neuromodulators on the intrinsic excitability of neurons. Neuromodulators can continuously modulate the firing pattern of a neuron by altering the conductances of specific ion channels~\citep{Levitan1979,Kaczmarek1987}. In conductance-based modelling, this action is typically studied by varying maximal conductance parameters of the targeted channels and analysing the corresponding change in neuronal activity. Such studies are impractical in existing integrate-and-fire models because the abstract tuning parameters lack physiological interpretation. As a consequence, two different firing patterns might require very different sets of parameters in the abstract model. Furthermore a continuous modulation of the abstract parameters is difficult to relate to a continuous modulation of maximal conductances. 

The proposed Multi-Quadratic Integrate-and-Fire (MQIF) model has few tuning parameters as well, but each parameter has a specific physiological interpretation that enables a mapping from the abstract parameter space to the space of maximal conductances. The classical QIF model uses a quadratic current to capture that excitability involves a region of negative conductance in the instantaneous I-V curve. The minimum of the quadratic is a balance point where the local conductance changes sign. Our fundamental ansatz is that such a balance might occur in distinct timescales and that the modulation of excitability primarily rests on a modulation of those distinct balances between negative and positive conductances. In the MQIF model, each balance is modelled through a quadratic current, and each timescale is modelled with a distinct first-order linear filter. Each quadratic current only requires two parameters: the location of its minimum and its curvature. Those parameters have direct physiological interpretation and their modulation captures with surprising accuracy important modulation properties of neurons, such as the continuous transition between tonic spiking and bursting, or between Type I and Type II excitability.

\section{Integrate-and-fire realisations of conductance-based models}
\label{sec:integr-fire-real}

All integrate-and-fire models and conductance-based models share the scalar differential equation
\begin{equation}
  C \dot{V} = - I_V + I
  \label{eq:neuron-model}
\end{equation}
that models a neuron as a capacitive electrical circuit obeying Kirchhoff's current law: the capacitive current $C \dot{V}$ is equal to the sum of the external current $I$ and the intrinsic current $-I_V$.

In the simplest Linear Integrate-and-Fire (LIF) model~\citep{Hill1936,Stein1965,Knight1972}, the intrinsic current is a resistive ohmic current: $I_V(V) = g_L(V-V_L)$ where $g_L$ is the leak resistance and $V=V_L$ the equilibrium potential in the absence of external input current. The graph of $I_V(V)$ is the static characteristic of the electrical circuit. In experimental electrophysiology, it is called the I-V curve.

In conductance-based models, the intrinsic current is the sum of several ohmic currents with conductances that are modulated by gating variables: ions flow through the cellular membrane through ion-specific channels with given voltage dependence and given kinetics. Conductance-based models are rooted in the seminal work of Hodgkin and Huxley~\citep{Hodgkin1952}, who explained the fundamental biophysical mechanism of excitability by decomposing the voltage-gated current as the sum of a fast inward current and a slow outward current. Assuming that the fast current is instantaneous, a qualitative description of their model which is sufficient for the purpose of this paper is as follows:
\begin{align}
C \dot{V} &=  -I_f(V) -I_s(V_s) + I \label{eq:HH-qualitative}\\
\tau_s \dot{V_s} &= V - V_s \nonumber
\end{align}
where the assumption $ C \ll \tau_s$ ensures a fast-slow timescale separation. The \emph{slow} variable $V_s$ has the interpretation of the \emph{fast} voltage filtered through a low-pass filter. 
The choice $I_f(V)=\frac{V^3}{3}-k_f V$ and $I_s(V_s)=k_s V_s$ leads to FitzHugh-Nagumo model~\citep{FitzHugh1961,Nagumo1962}, the popular two-dimensional qualitative reduction of Hodgkin-Huxley model. For $k_s>k_f>0$, the fast-slow model captures the fundamental property of excitability: the static I-V curve $I_f(V)+I_s(V)$ is monotonic, i.e. purely resistive, but the fast I-V curve $I_f(V)$ is hysteretic, i.e. has a local region of negative resistance. 

\begin{figure}[!h]
  \centering
  \includetikz{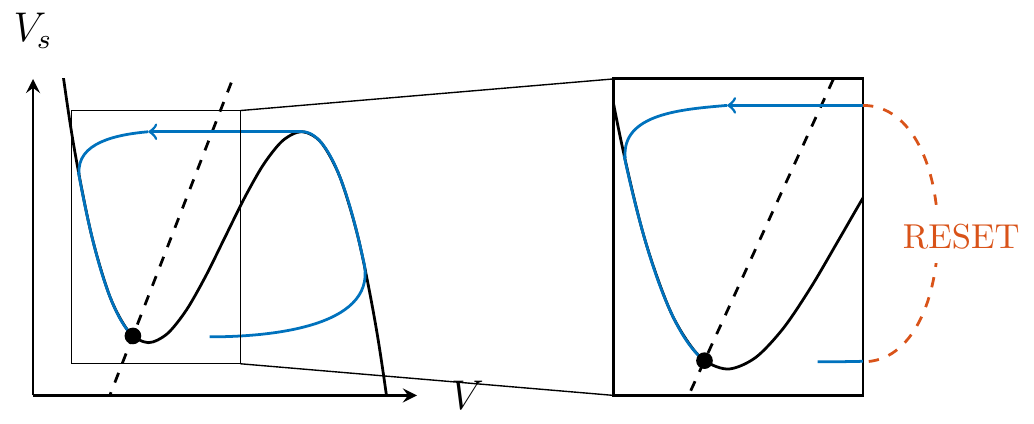}
  \caption{Phase portrait of FitzHugh-Nagumo model (left). Magnification of the region modelled by the Izhikevich model with reset (right). The $V$- and $V_s$-nullclines are drawn as full and dashed lines, respectively, and the stable fixed point as a filled circle.}
  \label{fig:FN-IF}
\end{figure}

The planar model Eq~(\ref{eq:HH-qualitative}) is a continuous-time model that captures the qualitative properties of a spike and that has a direct physiological interpretation. In particular, the fast and slow I-V curves can be identified from voltage-clamp experimental data or from a detailed computational conductance-based model, see~\citep{Drion2015}. However, the simulation of the model is not economical, because the model is nonlinear and stiff due to the timescale separation. The simulation can be made economical by acknowledging the all-or-none nature of the spike, replacing the large excursion in the phase plane by an instantaneous reset mechanism. The continuous-time model simulation is only needed to capture the subthreshold dynamics, which only requires the knowledge of the model in the neighbourhood of the equilibrium potential as illustrated in Figure~\ref{fig:FN-IF}. This simplification leads to an integrate-and-fire model.

The integrate-and-fire model has physiological interpretation because it satisfies the following two constraints: (i) the expression of the intrinsic current $I_V$ provides a qualitative description of the conductance-based model in the range of continuous-time simulation, i.e. $V \le V_c$ with $V_c$ the cut-off voltage, and (ii) the reset rule corresponds to a shortcut description of the corresponding continuous-time model trajectory. 

Popular integrate-and-fire models of the literature have this property: for instance, the model of Izhikevich~\citep{Izhikevich2003} adopts the quadratic-linear description $I_f(V) = -0.04 V^2 -5 V$, $I_s(V_s) = V_s$, whereas the model of Brette and Gerstner~\citep{Brette2005} adopts the exponential-linear description $I_f(V)= g_L (V-V_L)- g_L \Delta_T \exp(\frac{V-V_T}{\Delta_T})$, $I_s(V_s) = V_s$. Both models have the same local normal form. In particular, both the quadratic and exponential currents capture the fold bifurcation that organises the spike upstroke. The model of Izhikevich takes the general form
\begin{align}
  & & \text{if }V &\ge 30: \nonumber\\
  \dot{V} &= 0.04V^2 + 5V + 140 - V_s + I & V &\gets c \label{eq:Izhikevich}\\
  \dot{V_s} &= a(bV-V_s) & V_s &\gets V_s + d. \nonumber
\end{align}
and has been shown to replicate a broad range of firing activity for different choices of its parameters~\citep{Izhikevich2003,Izhikevich2007}. It certainly provides an economical simulation model of the planar model Eq~(\ref{eq:HH-qualitative}) for the choice of parameters $a=0.02$ \& $b=0.2$ and for a reset rule that satisfies the physiological constraint, e.g. $c=-65$ mV \& $d=2$. It also connects to the classical literature on integrate-and-fire models: it can be interpreted as the classical Quadratic Integrate-and-Fire (QIF) model augmented with one adaptation variable that models the refractoriness of excitability.  

The quadratic description of the intrinsic current in Eq~(\ref{eq:Izhikevich}) is thus an important connection between integrate-and-fire modelling and conductance-based modelling. Mathematically, it provides a local normal form of the fast I-V curve near threshold and captures the fold bifurcation that organises the spike upstroke in the fast timescale. Physiologically, it captures the local negative conductance of the circuit brought by the fast activation of an inward (sodium) current. The localisation of this negative conductance in time (fast) and amplitude (near the resting potential) is the fundamental property of excitability captured both by the planar model Eq~(\ref{eq:HH-qualitative}) and the integrate-and-fire model Eq~(\ref{eq:Izhikevich}).

\section{Integrate-and-fire modelling of fast and slow regenerativity}
\label{sec:fast-slow-regen}

\begin{figure}[!h]
  \centering
  \includetikz{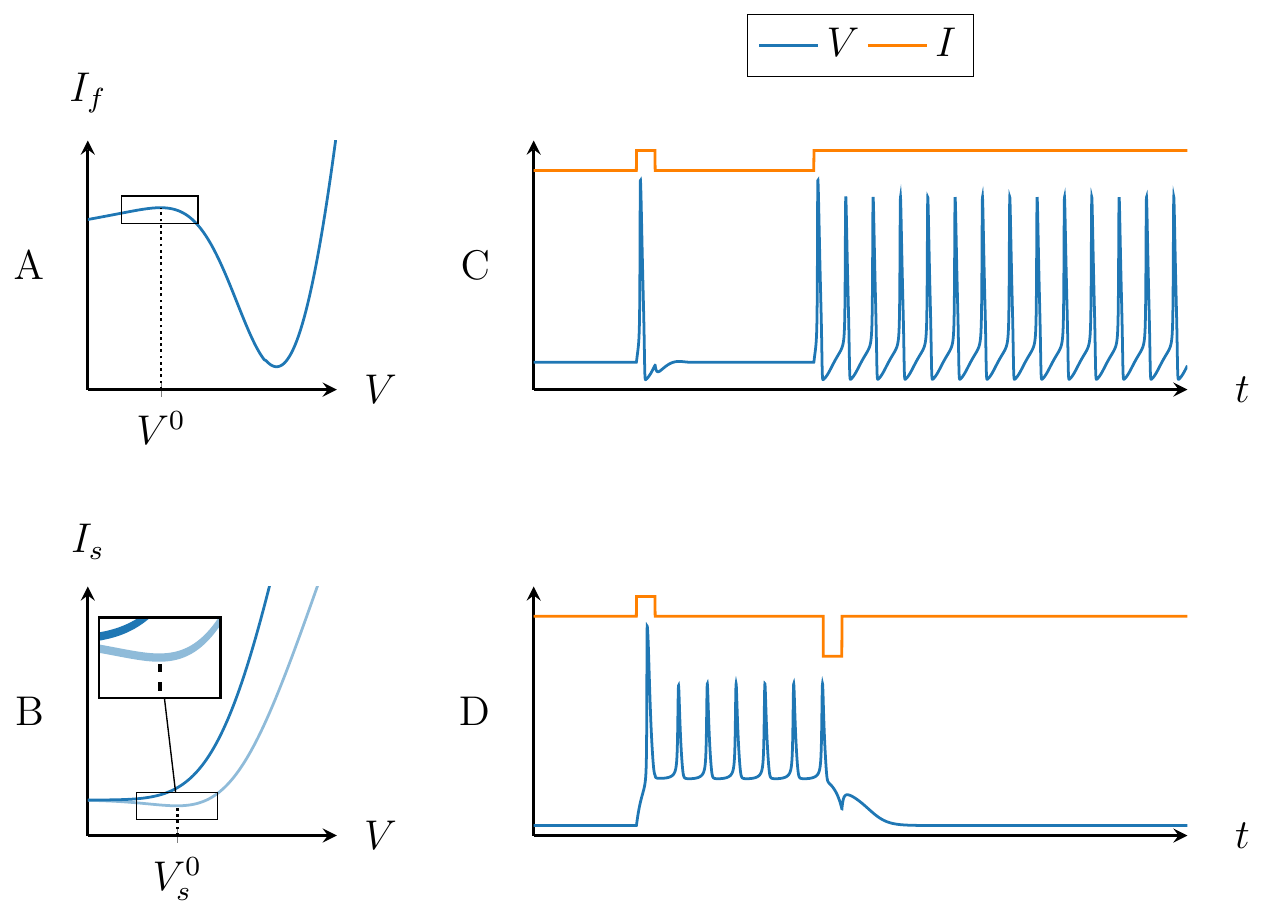}
  \caption{Fast (A) and slow (B) current in the Hodgkin-Huxley model with $V_K = -12$ mV (dark blue) and $V_K = 10$ mV (light blue). $V^0$ and $V_s^0$ are the points of balance of regenerativity and restorativity in the fast and slow timescale respectively. Corresponding voltage traces for the monostable model with $V_K = -12$ mV (C) and bistable model with $V_K = 10$ mV (D).}
  \label{fig:HH-VK}
\end{figure}

Starting with the work of Hodgkin and Huxley, conductance-based modelling proved extremely efficient at modelling not only the onset of an action potential but any type of electrical activity recorded in neurons. We choose to illustrate this generalisation with an insightful example grounded in the experimental work of Moore~\citep{Moore1959} published just a few years after the Hodgkin-Huxley model and subsequently modelled by Rinzel in 1985~\citep{Rinzel1985}. Moore observed that the slow I-V curve of the squid axon studied by Hodgkin and Huxley could be made non-monotonic by changing the extracellular concentration of potassium. Rinzel analysed this experiment by studying the effect of varying the corresponding parameter in Hodgkin and Huxley model (i.e. the Nernst potential $V_K$) and showed that it resulted in a bistable behaviour between rest and spike, a behaviour not observed in the nominal model. The simulation is reproduced in Figure~\ref{fig:HH-VK}. It shows that a qualitatively novel behaviour occurs in the fundamental model of excitability by slightly deforming the slow I-V curve of the model from monotonic to non-monotonic.

The region of negative conductance in the slow I-V curve has a direct physiological interpretation: the potassium current is an outward current for $V > V_K$ but an inward current for $V < V_K$. By manipulating the parameter $V_K$, the potassium current might be changed from outward to inward in the vicinity of the resting potential, turning the activation of potassium channels from a source of slow positive conductance to a source of slow negative conductance.

Manipulating the Nernst potential $V_K$ is not a physiological modulation of a neuronal behaviour, but many other channels not included in the original model of Hodgkin and Huxley do have the ability to shape the slow current $I_s(V_s)$ with a source of negative conductance. Such channels are called \emph{slow regenerative}. They include many calcium channels and some fast potassium channels, see~\citep{Hille2001}. In the same way as fast regenerative channels shape the negative conductance of the fast I-V curve, slow regenerative channels shape the negative conductance of the slow I-V curve.

There is ample evidence that slow regenerative channels contribute to neuronal activity in an essential manner. In the same way as non-monotonicity in the fast I-V curve is the fundamental signature of spike excitability, i.e. \emph{fast} excitability, non-monotonicity in the slow I-V curve is the fundamental signature of \emph{slow} excitability. It governs important phenomena such as spike latency, afterdepolarisation potential, and bursting, to cite a few.

\begin{figure}[!h]
  \centering
  \includetikz{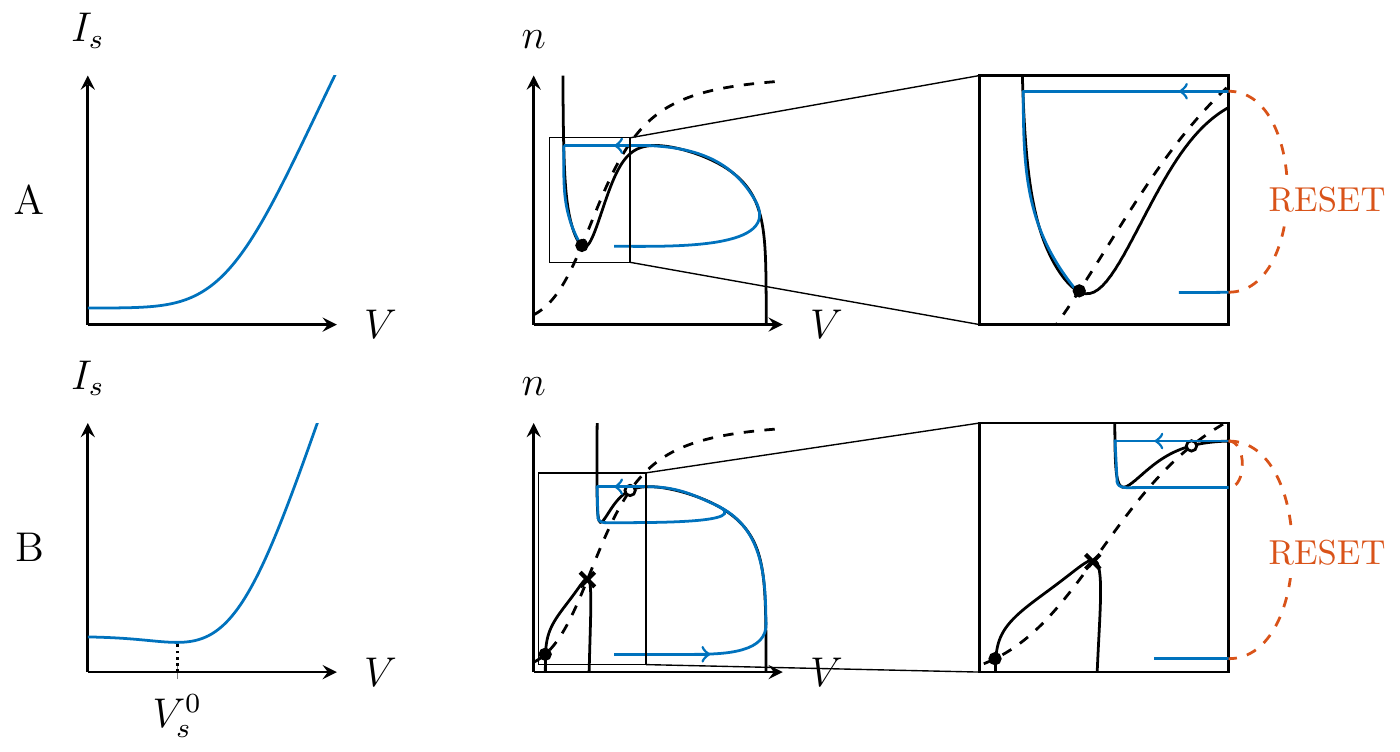}
  \caption{Comparison between the Hodgkin-Huxley model with $V_K = -12$ mV (A) and $V_K = 10$ mV (B). Left: slow I-V curves; middle: reduced models (see Methods); right: corresponding integrate-and-fire models. In the phase portraits, $V$- and $n$-nullclines are drawn as full and dashed lines, respectively. The stable and unstable fixed points are represented by filled and open circles, respectively, and the saddle point by a cross.}
  \label{fig:HH-PP}
\end{figure}

The phase portrait of a model with non-monotonicity both in the fast and the slow I-V curves was first studied in~\citep{Drion2012}, see also~\citep{Franci2012}. It was shown that the subthreshold dynamics are well described by a quadratic-quadratic description of the I-V curves, leading to the Multi-Quadratic Integrate-and-Fire (MQIF) model:
\begin{align}
  & & \text{if }V &\ge V_{max}: \nonumber\\
  C\dot{V} &= \bar{g}_f(V-V^0)^2-\bar{g}_s(V_s-V_s^0)^2+I \label{eq:MQIF} & V &\gets V_{r}\\
  \tau_s\dot{V}_s &= V-V_s & V_s &\gets V_{s,r} \nonumber.
\end{align}
Figure~\ref{fig:HH-PP} illustrates the phase portraits of both the reduced Hodgkin-Huxley model (see Methods) and an integrate-and-fire approximation in the same scenario as in Figure~\ref{fig:HH-VK}.

\begin{figure}[!h]
  \centering
  \includetikz{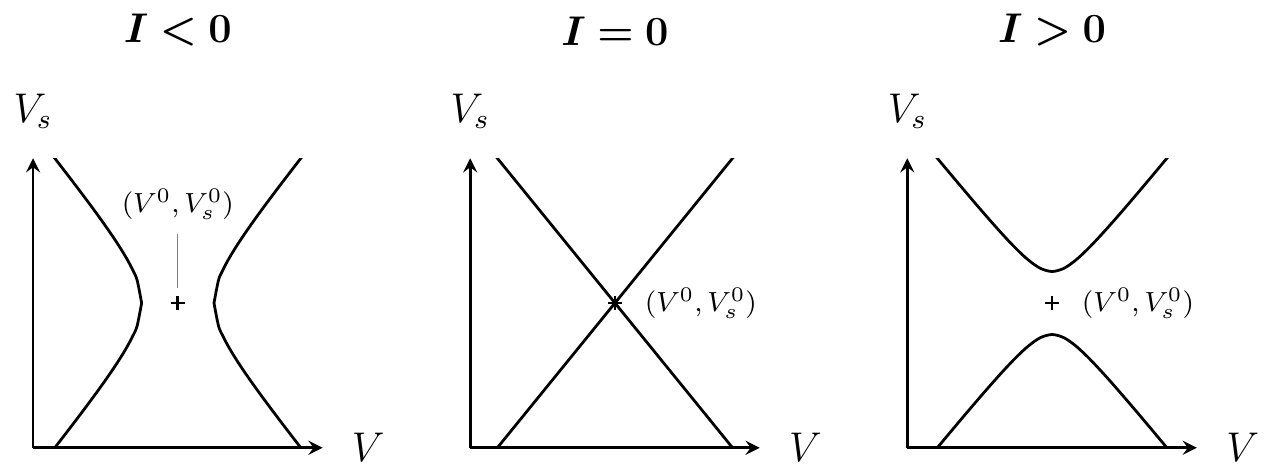}
  \caption{$V$-nullclines of the MQIF model for different input currents $I$. The parameters $V^0$ and $V_s^0$ move the nullcline horizontally and vertically, repectively.}
  \label{fig:MQIF-PP}
\end{figure}

The quadratic function in $V_s$ in Eq~(\ref{eq:MQIF}) strongly impacts the phase portrait, causing the mirroring of the quadratic $V$-nullcline (Figure~\ref{fig:HH-PP}B, right). The parameter $V_s^0$ is a critical parameter of the model. For $V_s^0 < V^0$, the phase portrait is very similar to classical QIF model reviewed in the previous section, see Figure~\ref{fig:HH-PP}A (right). In contrast, for $V_s^0 > V^0$, the stable equilibrium is on the lower branch of the $V$-nullcline (Figure~\ref{fig:HH-PP}B, right). This model is bistable, with a saddle point separating the stable fixed point and the limit cycle spiking state. For  $V_s^0 < V^0$, the model is slow restorative near the resting potential, that is, the slow conductance is positive. In contrast, for  $V_s^0 > V^0$, the model is slow regenerative near the resting potential, that is, the slow conductance is negative. In other words, the parameter $V_s^0$ has the physiological interpretation of the balance between slow restorative and slow regenerative channels near the resting potential. This balance plays a critical role in the regulation of excitability, see e.g.~\citep{Franci2013}. The effect of the input current $I$ on the $V$-nullcline is illustrated in Figure~\ref{fig:MQIF-PP}.

Slow regenerativity is difficult to capture in the model Eq~(\ref{eq:Izhikevich}) or any other model that imposes a single quadratic $V$-nullcline. As will be shown in the next sections, simulating  dynamical phenomena induced by slow regenerativity in the model Eq~(\ref{eq:Izhikevich}) requires to change the sign of either  parameter $b$ or parameter $d$. Such parameter changes destroy the interpretation of the hybrid model as a local approximation of a reduced conductance-based model completed by a reset rule compatible with the global continuous-time dynamics. The resulting model moves away from the physiological interpretation and introduces undesirable artefacts. In particular, modifying the reset rule makes it impossible to replace the reset rule by an equivalent continuous-time phase portrait. As a consequence, the connection to conductance-based modelling is lost.  

In the rest of the paper, we will analyse in more detail the robust features of an integrate-and-fire model that models both fast and slow regenerativity, through a quadratic-quadratic description of the I-V curves, and the difficulty of capturing the same features in a model that only accounts for fast regenerativity, through a quadratic-linear description of the I-V curves.

\section{Robust features of slow regenerativity}
\label{sec:robust-features-slow}

Slow regenerative channels control slow excitability, which manifests itself in a number of electrophysiological properties beyond bistability: spike latency, afterdepolarisation potential (ADP), bursting, and slow spiking. We will now show that those properties are easily reproduced in the MQIF model because the model captures slow regenerativity, that is, allows for a non-monotonic I-V curve in both the fast and slow timescales. In contrast, we will show that capturing those properties in a model that does not capture slow regenerativity necessarily requires mathematical manipulations that move away from physiological interpretation and modulation capabilities.

\subsection{Bistability}
\label{sec:bistability}

\begin{figure}[!h]
  \centering
  \includetikz{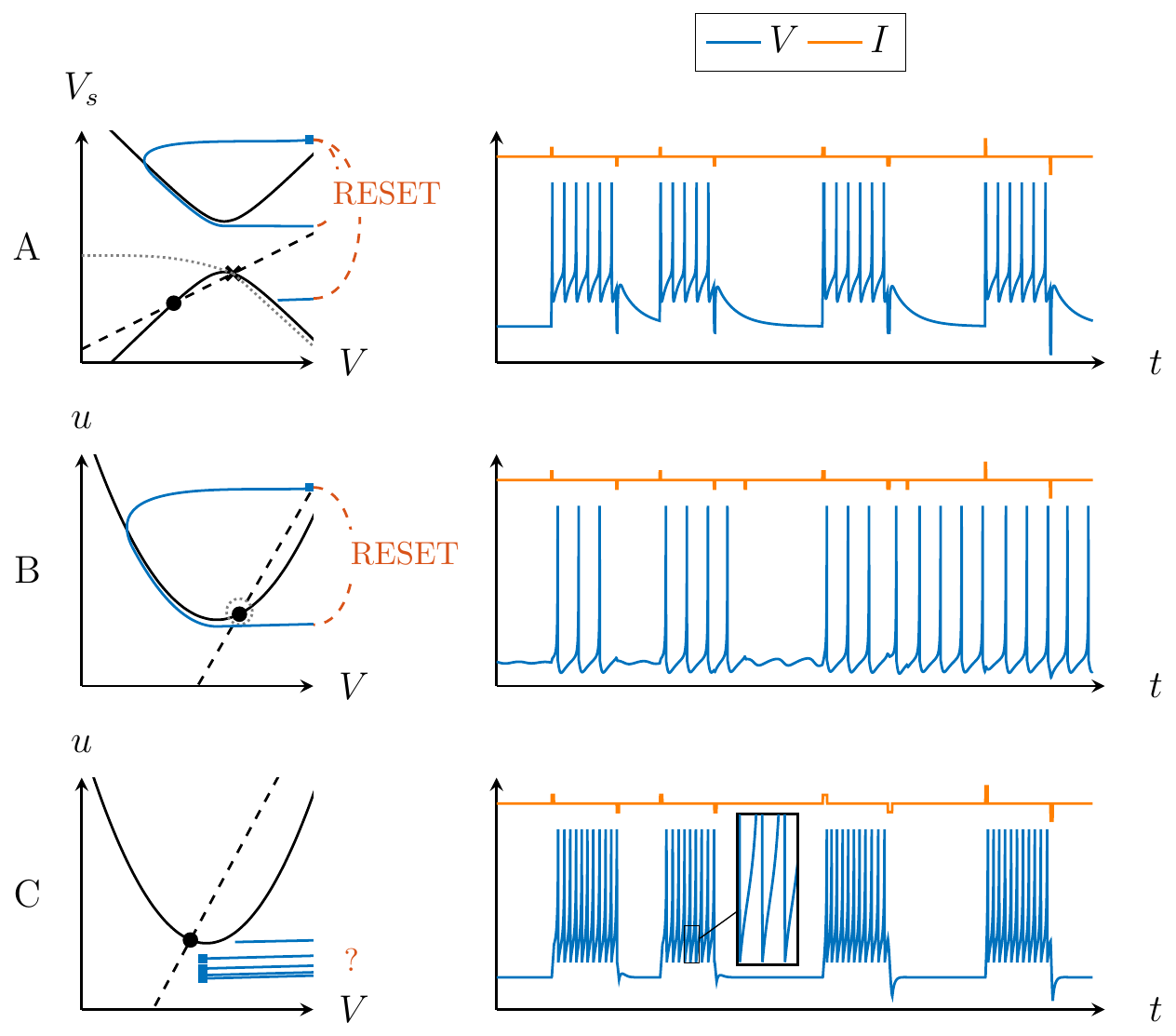}
  \caption{Bistability in the MQIF and Izhikevich models. Phase portraits (left) and current pulse responses (right) show robust and physiological bistability in the MQIF model (top). In a model that lacks slow regenerativity, bistability is either fragile to timescale separation (middle) or requires abstract reset rules (bottom) that lack physiological interpretation and modulation capabilities. In the phase portraits, fast ($V$) and slow ($V_s$ or $u$) nullclines are drawn as full and dashed lines, respectively. The stable fixed points are represented by filled circles, the saddle points by crosses and the reset points by blue squares. The boundary of the region of attraction of the stable fixed point is indicated by the grey dotted line.}
  \label{fig:bistability}
\end{figure}

In the previous section, robust bistability was shown to be a distinctive property of the MQIF model. When the nullclines intersect on the bottom branch of the $V$-nullcline, a saddle point separates a stable resting state and a stable spiking state (Figure~\ref{fig:bistability}A, left). This saddle point is persistent: it exists regardless of the timescale separation, and its stable manifold (grey dotted line) separates two robust basins of attraction of the two stable states~\citep{Franci2013}. Therefore a sufficiently large current pulse brings the model from spiking to resting or vice versa.
Figure~\ref{fig:bistability}A illustrates that switching between the two states is robust to timing, width and amplitude changes of the current input. Furthermore, the nature of the spiking exhibits properties seen in physiological recordings: afterhyperpolarisation (AHP) between the spikes and afterdepolarisation (ADP) before returning to the resting state.

The robust bistability exhibited in the MQIF model requires slow regenerativity. Figure~\ref{fig:bistability} illustrates that simulating bistability in a model that lacks slow regenerativity is either fragile or non-physiological. The simulation shown in Figure~\ref{fig:bistability}B is compatible with the classical phase portrait of excitability: the bistability results from a subcritical Hopf bifurcation. However, the basin of attraction of the stable fixed point (inside the grey dotted line) is necessarily small in that phase portrait, and decreases in size with increasing timescale separation. Therefore the switching from spiking to resting requires precise timing and amplitude of the current pulse. This non-persistent form of bistability is fragile~\citep{Franci2013}.

Figure~\ref{fig:bistability}C illustrates that the only way to make bistability robust in a model with timescale separation but without slow regenerativity is through a drastic change of the reset rule. The reset point is now below instead of above the $V$-nullcline. It approximately mimics what happens in the phase portrait of Figure~\ref{fig:bistability}A but the reset no longer corresponds to a continuous-time trajectory of the phase portrait in Figure~\ref{fig:bistability}C. The new reset rule recovers the robustness of bistability to the timing and amplitude of the current pulses, as illustrated on the right of Figure~\ref{fig:bistability}C, but the connection to conductance-based modelling is lost. The model no longer captures the continuous-time subthreshold dynamics that govern the electrical properties of the neuron between successive spikes.

\subsection{Spike latency}
\label{sec:spike-latency}

Another physiologically important manifestation of slow regenerativity is \emph{spike latency}. A variable latency preceding the first action potential in response to a current stimulus is observed in many electrophysiological recordings. Recent studies have explored the coding property of this particular behaviour for stimulus recognition in sensory systems~\citep{Chase2007,Gollisch2008,Zohar2011,Storchi2012}. Those studies suggest that it is an important and modulated quantity.

\begin{figure}[!h]
  \centering
  \includetikz{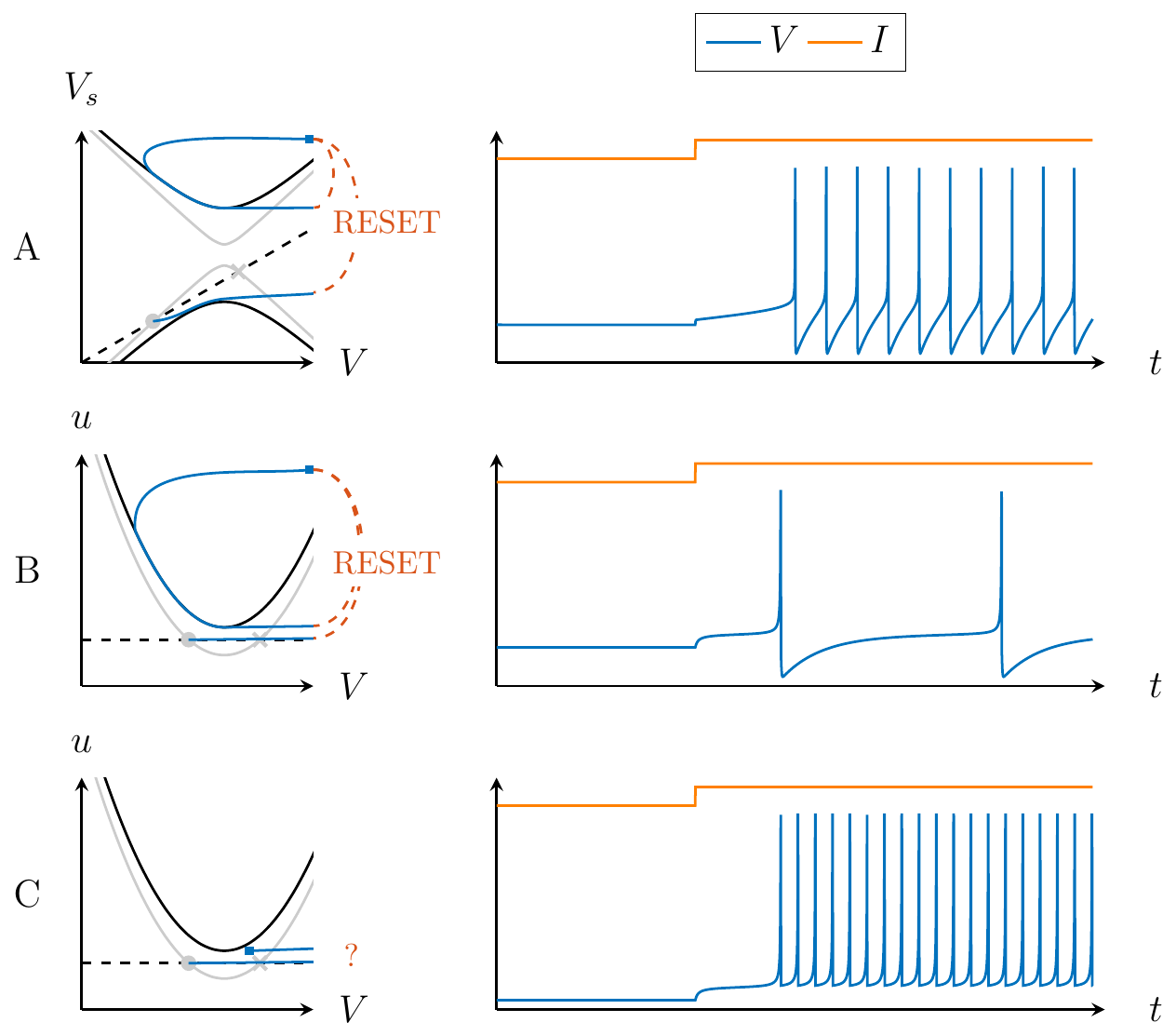}
  \caption{Spike latency in the MQIF and Izhikevich models. Phase portraits (left) and current step responses (right) show the presence of spike latency in the the MQIF model (top) and the Izhikevich model (middle and bottom). Latency of only the first spike requires to change the reset rule in a way that destroys the simple connection to continuous-time models (bottom). In the phase portraits, fast ($V$) and slow ($V_s$ or $u$) nullclines are drawn as full and dashed lines, respectively. The stable fixed points are represented by filled circles, the saddle points by crosses and the reset points by blue squares. The $V$-nullclines are drawn in grey and black, representing the low and high current phases of the simulation, respectively.}
  \label{fig:spike-latency}
\end{figure}

Spike latency is an inherent property of slow regenerativity. It is often associated to the transient potassium current $I_A$, which is slow regenerative because of its slow inactivation (see e.g.~\citep{Drion2015b}). Spike latency is well captured by the MQIF model as shown in Figure~\ref{fig:spike-latency}A. When the nullclines intersect on the bottom branch of the $V$-nullcline, a current pulse causes the trajectory to be attracted first by the ghost of the saddle-node bifurcation before moving to the spiking regime in the upper part of the phase portrait \citep{Franci2012}. From a dynamical systems viewpoint, the spike latency is nothing but the transient attractivity of a saddle point separating the two stable attractors. The transient is slow provided that the system exhibits slow regenerativity~\citep[Chapter 8]{Trotta2013}.

It is possible to obtain spike latency in a model without slow regenerativity by having an (almost) flat slow nullcline (see the discussion in~\citep[Section 7.2.9]{Izhikevich2007}). This ensures that the first spike will be attracted by the ghost of the saddle-node bifurcation and thus cause latency. Figure~\ref{fig:spike-latency}B illustrates the limitation of this mechanism: not only the first spike, but every spike is delayed by the saddle-node ghost. This method is fragile as well: a slightly higher slope of the slow nullcline destroys the spike latency \citep[Figure~17]{Franci2012}. It requires an unphysiological change of the reset rule to have the latency only on the first spike (Figure~\ref{fig:spike-latency}C).

\subsection{Afterdepolarisation (ADP)}
\label{sec:after-depol-potent}

Afterdepolarisations (ADP), also called depolarising afterpotentials (DAP), are distinctive depolarisations immediately following a spike, or a train of spikes, before returning to resting state. They are another distinctive signature of slow regenerativity, accompanying the transition from spiking to rest. The  particular signature of ADP has been associated to changes in excitability and bursting behaviour~\citep{Thompson1976,Metz2005,Yue2005}. More recently, ADPs have been linked to changes in interspike interval (ISI) variability~\citep{Fernandez2009} and plastic changes~\citep{Brown2009}.

\begin{figure}[!h]
  \centering
  \includetikz{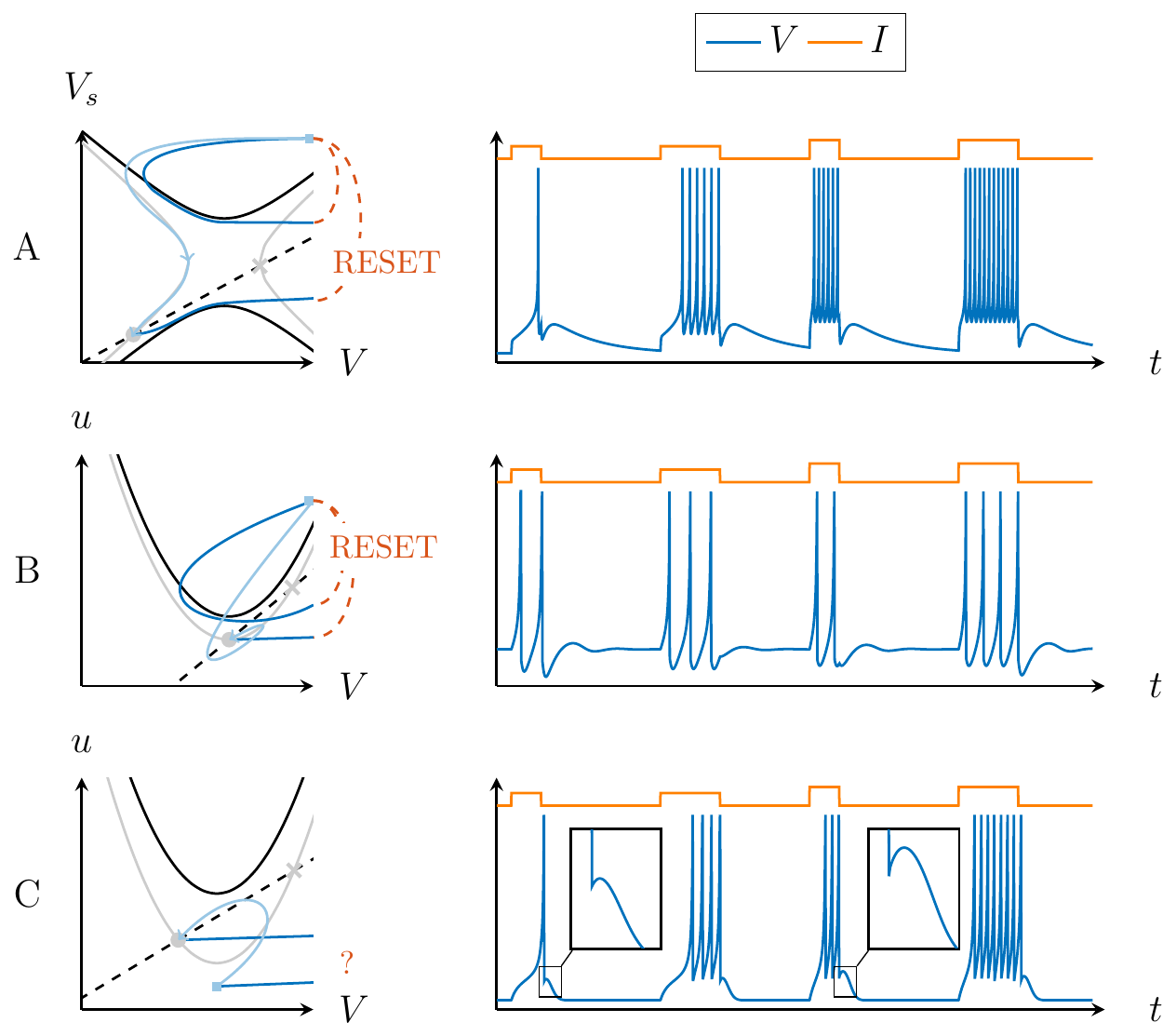}
  \caption{ADP in the MQIF and Izhikevich models. Phase portraits (left) and current pulse responses (right) show robust ADP in the MQIF model (top). The ADP in the Izhikevich model is either fragile (middle) or requires abstract reset rules (bottom) that lack physiological interpretation. In the phase portraits, fast ($V$) and slow ($V_s$ or $u$) nullclines are drawn as full and dashed lines, respectively. The stable fixed points are represented by filled circles, the saddle points by crosses and the reset points by blue squares. The grey $V$-nullclines and light blue trajectories represent the low current phase of the simulation, while the black $V$-nullclines and dark blue trajectories represent the high current phase.}
  \label{fig:ADP}
\end{figure}

In the MQIF model, the ADP is a necessary consequence of the hourglass-shaped $V$-nullcline when the baseline current is negative (solid black line in Figure~\ref{fig:ADP}A, left). A current step results in a delayed transition to spiking, see the light blue trajectories in Figure~\ref{fig:ADP}A (left). When the current step terminates, the trajectory returns to the stable resting state on the lower branch of the $V$-nullcline by closely following the hourglass shape of the $V$-nullcline. The ADP is robust and independent of the amplitude of the current step. In the language of~\citep{Franci2013}, the same singularity is responsible for the spike latency when switching from rest to spike and for the ADP when switching from spike to rest.

It is difficult to simulate an ADP in a model that lacks slow regenerativity, see for instance the discussion in~\citep[Section 7.3.5]{Izhikevich2007}. The mechanism is inherently fragile and disappears with increasing timescale separation. The fragility can be observed in Figure~\ref{fig:ADP}B, where the amplitude of the ADP depends strongly on the input. A similar result is shown in \citep[Figure~11]{Drion2012}.

Alternatively, an artificial ADP is created by changing the reset rule to a reset point under the $V$-nullcline. If a current step causes the stable fixed point to disappear, subsequent spikes will originate from below the $V$-nullcline. Once the current step has ended, the trajectory towards the fixed point will make an excursion towards the right as on the left of Figure~\ref{fig:ADP}C, resulting in an ADP. As mentioned before, resetting below the $V$-nullcline is unphysiological and eliminates the AHP and relative refractoriness. Finally, the magnitude of the ADP depends strongly on the timing and amplitude of the current steps, as illustrated on the right of Figure~\ref{fig:ADP}C. Figure~12 of \citep{Drion2012} shows a more detailed analysis of the fragility. This undesirable artefact is a consequence of the abstract reset rule.

\section{Bursting in integrate-and-fire models}
\label{sec:burst-integr-fire}

\subsection{Bursting in the MQIF model}
\label{sec:bursting-mqif}

Slow regenerativity is a central ingredient of bursting, because it is precisely the combination of fast \emph{and} slow regenerative channels that creates distinct mechanisms for the (fast) spike upstroke and the (slow) burst upstroke. The recent paper~\citep{Franci2017} illustrates that bursting is fragile and lacks modulation properties in the absence of slow regenerative channels. Likewise, we will now show that the same conclusion holds in integrate-and-fire models.

\begin{figure}[!h]
  \centering
  \includetikz{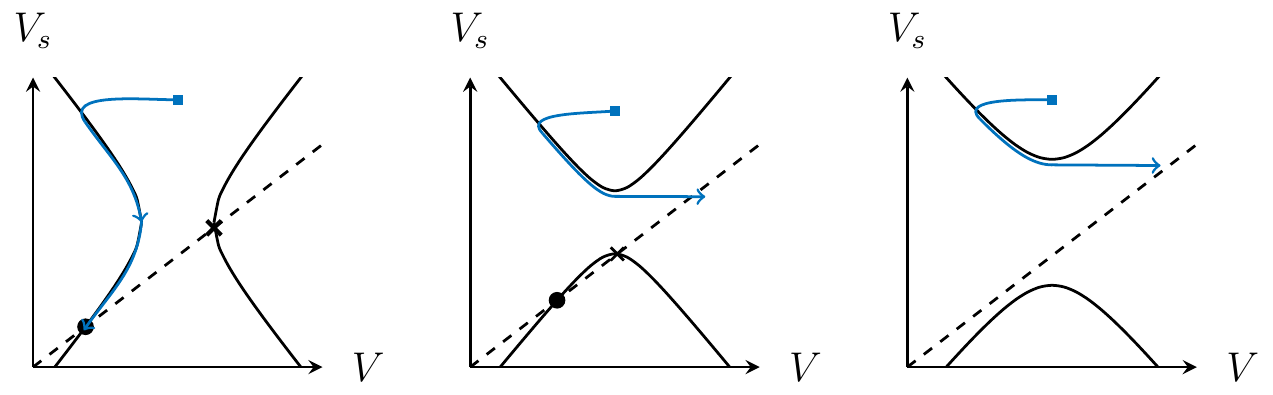}
  \caption{Phase portraits of the MQIF model with $V_s^0 > V^0$ for low (A) to high (C) current $I$. A: For low current the model has a single stable fixed point. B: For intermediate currents the model is bistable. C: At higher currents, only the spiking attractor persists. The $V$- and $V_s$-nullclines are drawn
as full and dashed lines, respectively. The stable fixed points are represented by filled circles, the saddle points by crosses and the reset points by blue squares.}
  \label{fig:PP-I}
\end{figure}

Bursting is the natural rhythm that emerges when adding adaptation to a bistable phase portrait such as the one in Figure~\ref{fig:bistability}A. Figure~\ref{fig:PP-I} shows the three phase portraits of the MQIF model when $V_s^0 > V^0$ and when the applied current is varied. For a low value of the applied current, the model has a single stable fixed point and a saddle point in the lower half-plane (Figure~\ref{fig:PP-I}, left). The stable fixed point is the only attractor. For intermediate values of the applied current, the model exhibits the bistable regime (Figure~\ref{fig:PP-I}, middle) where both spiking and rest are stable attractors. Depending on the initial condition, trajectories converge to one of the two attractors. For high values of the applied currents, the stable fixed point and saddle point disappear and only the spiking attractor persists (Figure~\ref{fig:PP-I}, right).

These three regimes provide the model with a hysteretic transition between rest and spiking, which is the fundamental mechanism of bursting. Ultraslow adaptation converts the hysteretic transition into a bursting oscillation. We introduce this \emph{ultraslow} timescale in the MQIF exactly as we introduced the \emph{slow} timescale: we add the new quadratic current $\bar{g}_{us}(V_{us}-V_{us}^0)^2$ where $V_{us}$ is now an ultraslow variable modelled as a filtered version of the voltage: 
\begin{align}
  & & \text{if }V &\ge V_{max}\nonumber:\\
  C\dot{V} &= \bar{g}_f(V-V^0)^2-\bar{g}_s(V_s-V_s^0)^2 & V &\gets V_{r}\nonumber\\
    &\quad -\bar{g}_{us}(V_{us}-V_{us}^0)^2+I \label{eq:MQIF_us}\\
  \tau_s\dot{V}_s &= V-V_s & V_s &\gets V_{s,r}\nonumber\\
  \tau_{us}\dot{V}_{us} &= V-V_{us} & V_{us} &\gets V_{us} + \Delta V_{us}.\nonumber
\end{align}
 
\begin{figure}[!h]
  \centering
  \includetikz{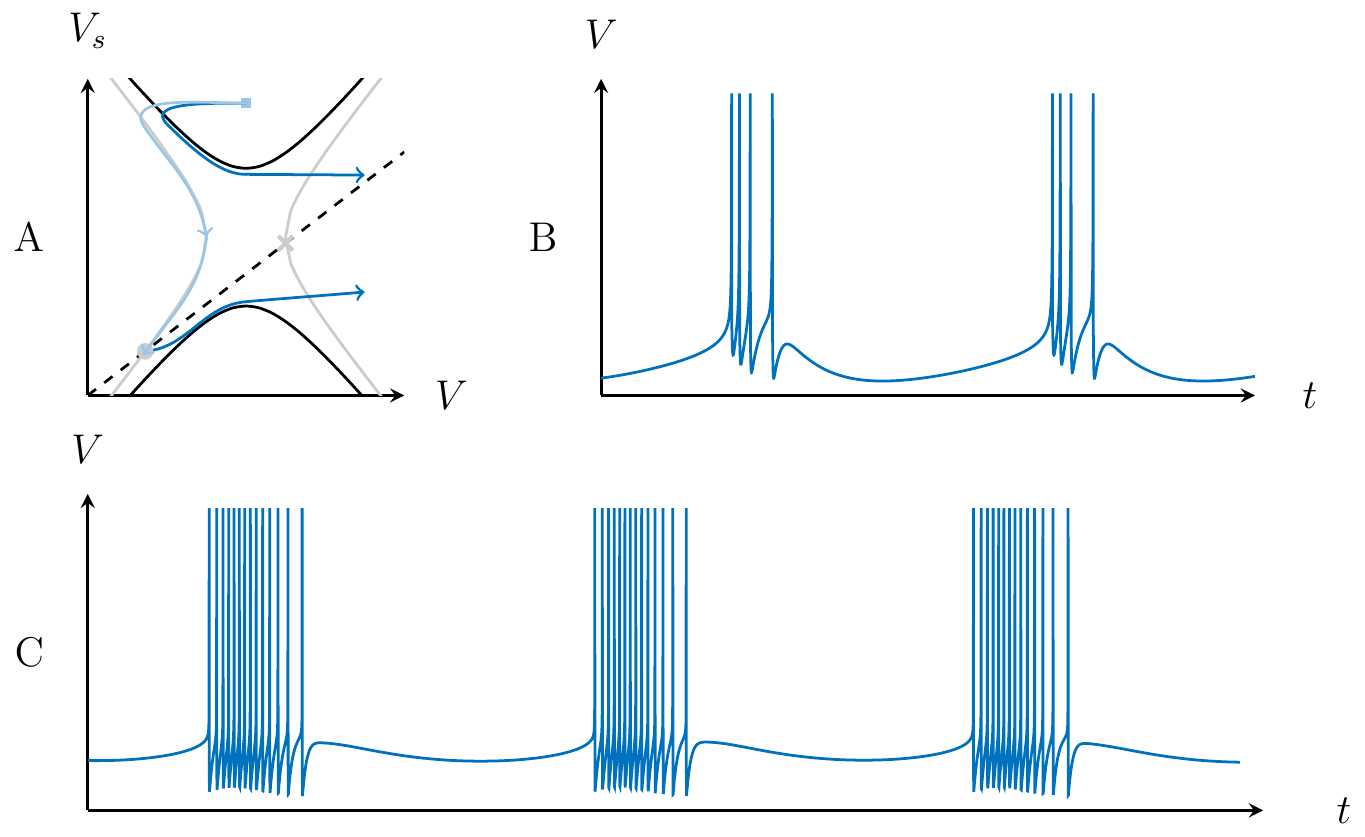}
  \caption{Bursting in the MQIF model. Phase portrait (A) and voltage traces of square-wave bursting (B) and parabolic bursting (C) in the MQIF model. The stable fixed point is represented a filled circle and the saddle point by a cross. The $V$- and $V_s$-nullclines are drawn
as full and dashed lines, respectively. The grey $V$-nullclines and light blue trajectories represent the silent phase of the burst, while the black $V$-nullclines and dark blue trajectories represent the active phase.}
  \label{fig:MQIF-bursting}
\end{figure}

The augmented model captures the balance of positive and negative conductances in three distinct timescales. The new parameter $V_{us}^0$ controls the balance between restorative and regenerative channels in the ultraslow timescale.  The role of this new balance is illustrated in Figure~\ref{fig:MQIF-bursting} which illustrates two distinct types of bursting classically referred to as \emph{square-wave} and \emph{parabolic} bursting.

Square-wave bursting (Figure~\ref{fig:MQIF-bursting}B) does not require ultraslow regenerativity. The ultraslow variable only provides the negative feedback adaptation that converts the hysteretic transition between tonic spiking and resting into a bursting oscillation. This ultraslow adaptation is achieved with a restorative ultraslow current, which corresponds to choosing  $V_{us}^0$ sufficiently low so that the ultraslow current is restorative over the whole voltage range of the bursting oscillation. 

Parabolic bursting (Figure~\ref{fig:MQIF-bursting}C), in contrast, requires ultraslow regenerativity, which is achieved in the MQIF model by increasing  the parameter $V_{us}^0$.  Regenerativity of the ultraslow current provides positive feedback in the ultraslow timescale, which accounts for  the increasing firing rate during the burst, the specific feature of parabolic bursting. The latency in the intraburst frequency is the ultraslow analogue of the spike latency studied in Section~\ref{sec:spike-latency}. The adaptation necessary to end the burst is moved to an even slower timescale, represented by the \emph{ultra-ultraslow} variable $V_{uus}$. This construction suggests that the construction of the MQIF model can be iterated through an arbitrary number of  timescales.  Previous models of parabolic bursting indeed rely on four distinct timescales, see ~\citep{Rinzel1987} or~\citep[Section 9.3.2]{Izhikevich2007}. The MQIF model captures parabolic bursting with the equations
\begin{align}
  & & \text{if }V &\ge V_{max}:\nonumber\\
  C\dot{V} &= \bar{g}_f(V-V^0)^2-\bar{g}_s(V_s-V_s^0)^2 & V &\gets V_{r}\nonumber\\
    &\quad -\bar{g}_{us}(V_{us}-V_{us}^0)^2 -\bar{g}_{uus}(V_{uus}-V_{uus}^0)^2+I \label{eq:MQIF_uus}\\
  \tau_s\dot{V}_s &= V-V_s & V_s &\gets V_{s,r}\nonumber\\
  \tau_{us}\dot{V}_{us} &= V-V_{us} & V_{us} &\gets V_{us} + \Delta V_{us}\nonumber\\
  \tau_{uus}\dot{V}_{uus} &= V-V_{uus} & V_{uus} &\gets V_{uus} + \Delta V_{uus}.\nonumber
\end{align}
with a restorative ultra-ultraslow timescale, achieved with a sufficiently low $V_{uus}^0$. 

\subsection{Bursting in the Izhikevich model}
\label{sec:burst-izhik-model}

\begin{figure}[!h]
  \centering
  \includetikz{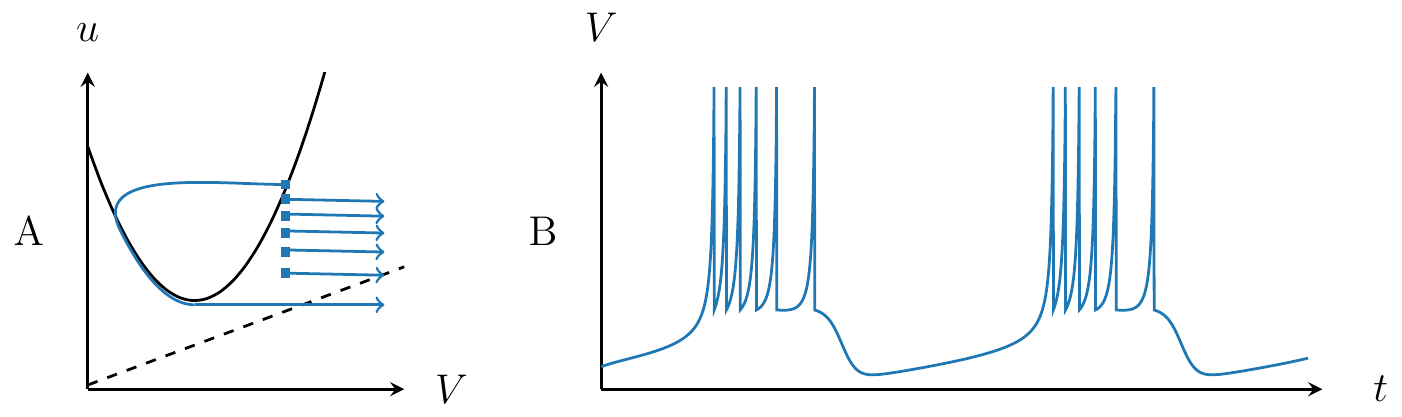}
  \caption{Bursting in the Izhikevich model. Phase portrait (A) and voltage trace (B) of bursting in the Izhikevich model. The $V$- and $V_s$-nullclines are drawn
as full and dashed lines, respectively. The reset points are represented by blue squares.}
  \label{fig:Izhikevich-bursting}
\end{figure}

Square-wave bursting in the Izhikevich model does not rely on a hysteresis loop switching between the rest and spiking state as in the MQIF. Instead, the reset point is moved to the right of the $V$-nullcline (see Figure~\ref{fig:Izhikevich-bursting}A). After several spikes, the slow variable $u$ will have increased sufficiently for the reset point to lie above the $V$-nullcline. This marks the end of the burst with a hyperpolarisation, after which the process starts over again.

Although this process captures the slow alternation of spiking and rest and the voltage trace looks like bursting, it lacks any connection to the physiology of bursting. As noted before, resetting below the $V$-nullcline is undesirable when wanting to keep the spiking physiologically plausible. The typical features of slow regenerativity (see Section~\ref{sec:robust-features-slow}) are absent as well. Lastly, without a hysteresis loop, this model of bursting is inherently more fragile than the MQIF model. 

\section{Modulation in the MQIF model}
\label{sec:modul-MQIF}

Recent studies with conductance-based modelling suggest that the balance of restorative and regenerative currents in the slow timescale is key to the modulation of a firing pattern~\citep{Franci2013,Franci2014,Drion2015,Drion2015a}. Neuromodulators can finely tune this balance by controlling the expression of specific ion channels.  
The structure of the MQIF model provides a simple way to qualitatively control this balance in each timescale with the two parameters that determine a quadratic current $\bar g (V-V^0)^2$ in a given timescale: the apex $V^0$ determines the location of the balance in the voltage range and the conductance $\bar g$ determines the strength of the conductance in the vicinity of the resting potential. We will illustrate this qualitative correspondence between the modulation of a conductance-based model and of the MQIF model in two examples that have been previously studied through conductance-based modelling.
 
\subsection{Continuous modulation between tonic spiking and bursting}
\label{sec:modul-bursting}

The continuous modulation between tonic spiking and bursting has been reported in many electrophysiological recordings and is often associated to distinct behaviours and brain states \citep{Harris-Warrick1991,Marder2012,Lee2012}.

\begin{figure}[!h]
  \centering
  \includetikz{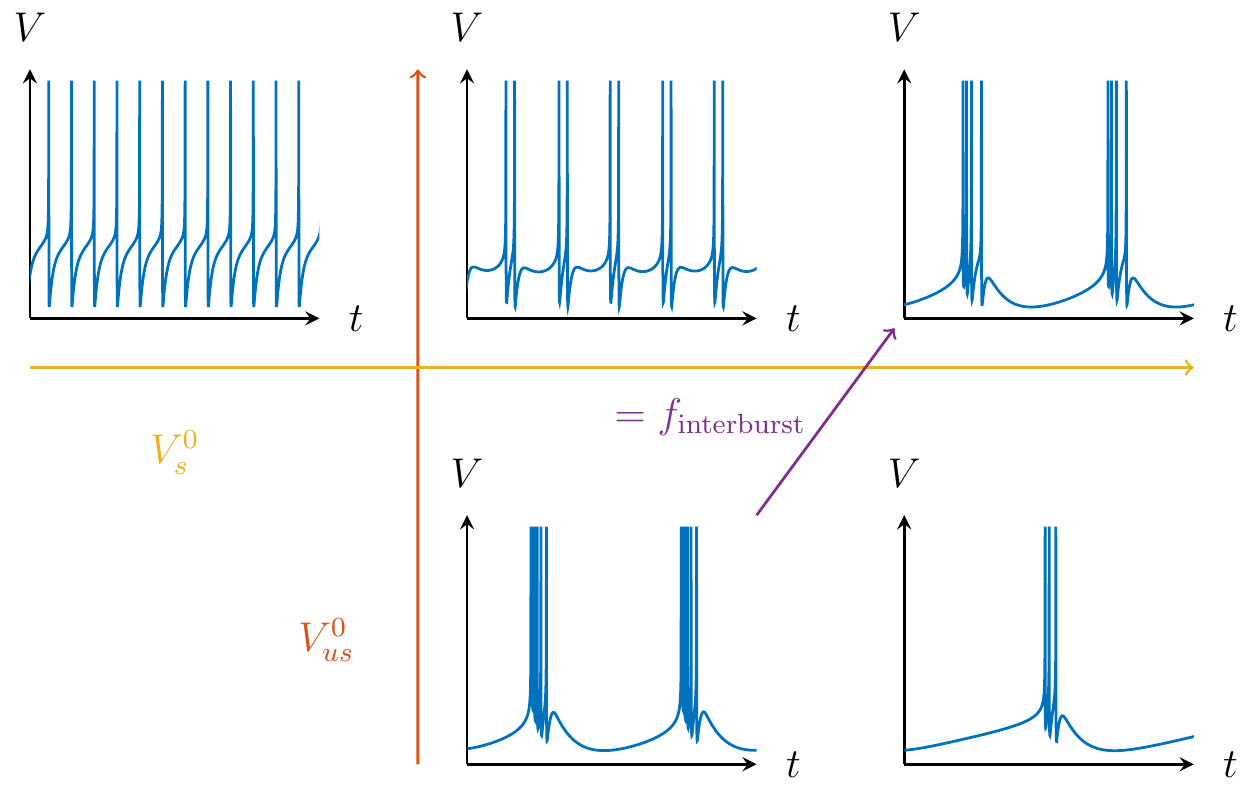}
  \caption{Modulating of bursting in the MQIF model. The MQIF model is modulated between tonic spiking and bursting by increasing $V_s^0$, changing the balance between slow restorativity and regenerativity (orange arrow). Decreasing $V_{us}^0$ increases the interburst frequency via an increase in ultraslow restorativity (red arrow). The interburst frequency can be kept constant while changing the intraburst frequency by simultaneously modulating $V_s^0$ and $V_{us}^0$ (purple arrow).}
  \label{fig:MQIF-mod}
\end{figure}

The recent paper~\citep{Franci2014} proposes that this modulation is the result of modulating the slow ionic current from restorative to regenerative. In the MQIF model, this modulation corresponds to modulate the slow balance parameter $V_s^0$ from a value smaller than the fast balance parameter $V^0$ to a value bigger than $V^0$. Figure~\ref{fig:MQIF-mod} (top) illustrates this modulation of $V_s^0$, which indeed results in a modulation from tonic spiking, via an increasing ADP to bursting, without detailed tuning of the model parameters. The modulation of the voltage activity closely resembles the temporal traces of physiological recordings \citep{Bal1993,Levitan1988}.

Similar to $V_s^0$, $V_{us}^0$ can be seen as the point of balance between ultraslow positive and negative feedback. The parameter is directly related to the expression of ion channels with ultraslow dynamics. As shown in Figure~\ref{fig:MQIF-mod}, an increase of $V_{us}^0$ decreases the ultraslow negative feedback and thus increases the number of spikes per burst and interburst frequency. By simultaneously modulating $V_s^0$ and $V_{us}^0$, the interburst frequency can be kept constant while changing the number of spikes per burst (purple arrow in Figure~\ref{fig:MQIF-mod}).

\subsection{Modulation between Type I and Type II excitability}
\label{sec:modul-excit-type}

\begin{figure}[!h]
  \centering
  \includetikz{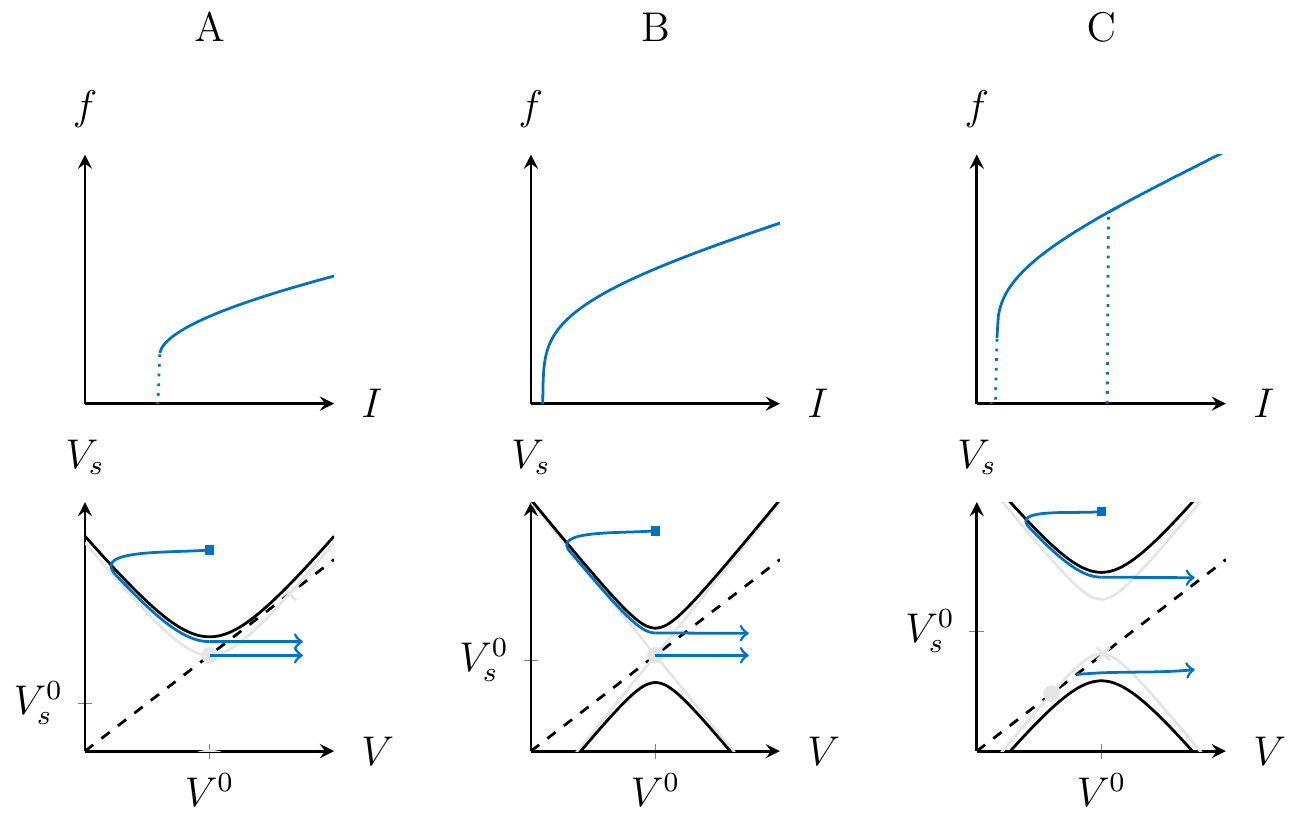}
  \caption{Modulation of the excitability type in the MQIF model. Changing $V_s^0$ in the MQIF model results in different types of excitability: Type II for $V_s^0 < V^0$ (A), Type I for $V_s^0 = V^0$ (B) and Type II*~\citep{Drion2015a} for $V_s^0 > V^0$ (C). The f-I curves (top) show the spike frequency as a function of the input current $I$. The phase portraits (bottom) show the $V$-nullcines just before (grey) and after the bifurcation (full black). The $V_s$-nullclines are drawn
as dashed lines. The stable fixed points are represented by filled circles, the saddle points by crosses and the reset points by blue squares.}
  \label{fig:MQIF-type}
\end{figure}

The recent paper~\citep{Drion2015a} revisits the classical distinction between Type I and Type II excitability. The paper proposes that the modulation of slow regenerativity is sufficient to  continuously modulate the firing pattern between the two types of excitability and uses this insight to introduce a third type of excitability called Type II*. The interpretation is that slow firing requires a very small conductance during the refractory period, which can be achieved by balancing slow restorative and a slow regenerative currents. The discussion in~\citep{Drion2015a} is with a classical conductance-based model. Here we reproduce the same modulation in Figure~\ref{fig:MQIF-type} with the MQIF model. Modulation between the three types of excitability is achieved very simply by modulating the slow balance parameter $V_s^0$. All simulations assume $g_s < g_f$, meaning that the asymptotic slope of the $V$-nullcline is greater than that of the $V_s$-nullcline.

When $V_s^0 < V^0$, there is no slow positive feedback (Figure~\ref{fig:MQIF-type}A). As the current $I$ increases, the fixed point loses its stability in a subcritical Hopf bifurcation. There is a sudden switch to spiking and an increase of current barely affects the spike frequency. This Type II excitability is observed in the f-I curve of Figure~\ref{fig:MQIF-type}A (top).

Another type of excitability is seen when  $V_s^0 = V^0$ (Figure~\ref{fig:MQIF-type}B). For $I=0$, the two branches of the $V$ nullcline collide and intersect the $V_s$-nullcline at the same point. The result is that the fixed point loses stability in a saddle-node on invariant circle (SNIC) bifurcation. This bifurcation leaves a ghost region through which subsequent spikes will pass and thus results in Type I excitability, whereby the spike frequency starts at 0 Hz and increases with the current. The f-I curve in Figure~\ref{fig:MQIF-type}B (top) clearly shows a continuous increase of the spike frequency starting from 0 Hz.

The novel Type II* excitability discussed in~\citep{Drion2015a} occurs in the situation $V_s^0 > V^0$, which is shown in Figure~\ref{fig:MQIF-type}C. Starting on the lower branch of the $V$-nullcline, the fixed point loses stability in a saddle-node bifurcation as $I$ increases. This slightly delays the first spike (see Section~\ref{sec:spike-latency} on spike latency), but subsequent spiking will have a finite frequency. This excitability resembles Type II excitability, but is different as it exhibits hysteresis. Decreasing $I$ again, the spiking only stops when the limit cycle on the upper branch disappears, which occurs at a lower current. The f-I curve in Figure~\ref{fig:MQIF-type}C (top) shows the hysteresis and discontinuity at both bifurcation points.

\section{Conclusion}
\label{sec:discussion}

The MQIF model discussed in this paper rests on a simple ansatz: the regulation of excitability rests on a balance between restorative and regenerative ionic currents in distinct timescales. The model captures each timescale with a distinct first-order filter and each balance with a quadratic current that models the balance locally around the resting potential. The action potential only requires one (fast) balance and is adequately modelled with one (fast) quadratic current. This has long been recognised through QIF modelling. Phenomena such as bistability, spike latency and afterdepolarisation potential (ADP) are all manifestations of a second balance in the slow timescale, which is modelled through a second (slow) quadratic current. Three timescale phenomena such as bursting motivate a third timescale and a third (ultraslow) quadratic current, etc.

When a quadratic current is regenerative near the resting potential, it provides \emph{positive} feedback. This positive feedback creates bistability in the corresponding timescale: bistability between low and high resting state potentials with one timescale, bistability between resting and spiking with two timescales, bistability between resting and bursting with three timescales, and so on. When a quadratic current is restorative near the resting potential, it provides \emph{negative} feedback or \emph{adaptation}: a stable equilibrium with one timescale, a stable limit cycle with two timescales, a stable bursting oscillation with three timescales, and so on. The MQIF model reproduces all such phenomena by tuning the balance independently in each timescale. The robustness of the construction owes to the timescale separation between the distinct first-order filters. 

The MQIF model is easily modulated between distinct types of excitability because of the specific interpretation of each parameter as a balance parameter or as a conductance parameter
in a specific timescale. This structure allows an easy mapping from physiological modulation studies on conductance-based models to qualitatively similar modulation studies in integrate-and-fire models. This is of considerable interest for network studies. Recent work suggests the important role of cellular slow regenerativity in the robustness and modulation properties at the network level~\citep{Dethier2015a}. So far, such studies have not been possible with large networks because they require detailed conductance-based modelling.

Simulation of the MQIF model is economical. This is because the linear filters can be integrated analytically, meaning that the simulation of one neuron never exceeds the numerical integration of a single nonlinear differential equation that entirely captures the continuous-time subthreshold dynamics.

This paper has illustrated that the MQIF model provides a \emph{qualitative} alternative to various modulation studies performed with conductance-based modelling. It will be interesting to devote further work to automate the tuning of the MQIF model for a given conductance-based model. Dynamic Input Conductances~\citep{Drion2015} suggest that this problem is tractable and that the important parameters of a MQIF model could be extracted either from a conductance model or even directly from electrophysiological recordings. Such results would revive the importance of integrate-and-fire modelling for neuromodulation computational studies.

\section{Methods}
\label{sec:materials-methods}

\subsection{Software}
\label{sec:software}

Simulations were performed with Brian 2~\citep{Goodman2009,Stimberg2014} using the model equations stated in the text. I-V curves and the phase portraits of the reduced Hodgkin-Huxley model were calculated with MATLAB. All figures were drawn using the \LaTeX packages PGF/TikZ and PGFPlots.

\subsection{Hodgkin-Huxley model and I-V curves}
\label{sec:hodgkin-huxley-model}

Figures for the Hodgkin-Huxley model use the original model equations as in~\citep{Hodgkin1952} with standard parameters ($C = 1$ \textmu F/cm\textsuperscript{2}, $E_L = 10.6$ mV, $E_K = -12$ mV, $E_{Na} = 115$ mV, $g_L = 0.3$ mS/cm\textsuperscript{2}, $g_K = 36$ mS/cm\textsuperscript{2}, $g_{Na} = 120$ mS/cm\textsuperscript{2}). For the bistable situation, $E_K$ was changed to 10 mV. The reduction to the $V$-$n$ was done as in~\citep[Section~5.2.1]{Izhikevich2007}: setting $m=m_{\infty}(V)$ and $h=(0.89-1.1n)$.

For Figure~\ref{fig:HH-VK}C, the baseline current $I$ equals 0 \textmu A/cm\textsuperscript{2}, with pulses of 10 \textmu A/cm\textsuperscript{2}. Figure~\ref{fig:HH-VK}D has a baseline current $I$ of $-10$ \textmu A/cm\textsuperscript{2} and pulse amplitudes of 10 \textmu A/cm\textsuperscript{2} and $-20$ \textmu A/cm\textsuperscript{2}.

The I-V curves of Figure~\ref{fig:HH-VK} and Figure~\ref{fig:HH-PP} are the integral over the voltage of the Dynamic Input Conductances, obtained from the conductance-based model as described in~\citep{Drion2015}. Specifically,
\begin{align}
  I_f(V') &=  \int_{E_K}^{V'} \left(C\frac{\partial \dot{V}}{\partial V} + g_f \right) dV\\
  I_s(V') &= \int_{E_K}^{V'} g_s dV,
\end{align}
with $g_f$ and $g_s$ as in~\citep{Drion2015}. For this method, $m$ was chosen as the fast variable, $h$ and $n$ as the slow variables.

\subsection{Integrate-and-fire models simulations}
\label{sec:if-simul}

The simulations in Figure~\ref{fig:bistability}-\ref{fig:ADP}A use Eq~\eqref{eq:MQIF} with the parameters in Table~\ref{tab:MQIF} and the common parameters $C = 1$ ms, $\tau_s = 10$ ms, $V_r = -40$ and $V_{s,r} = -30$.

\begin{table}[!ht]
\centering
\caption{
{\bf Parameter values for simulations of the MQIF model.}}
\begin{tabular}{|l+c|c|c|c|}
\hline
 & {\bf $V^0$} & {\bf $V_s^0$} & {\bf $\bar{g}_f$} & {\bf $\bar{g}_s$}\\ \thickhline
  Figure~\ref{fig:bistability}A & $-40$ & $-35$ & 1 & 0.2\\ \hline
  Figure~\ref{fig:spike-latency}A & $-40$ & $-35$ & 1 & 0.5\\ \hline
  Figure~\ref{fig:ADP}A & $-40$ & $-39$ & 1 & 0.5\\ \hline
\end{tabular}
\label{tab:MQIF}
\end{table}

The simulations in Figure~\ref{fig:bistability}-\ref{fig:ADP}B \& Figure~\ref{fig:bistability}-\ref{fig:ADP}C use Eq~\eqref{eq:Izhikevich} with the parameters in Table~\ref{tab:Izhikevich}.

\begin{table}[!ht]
\centering
\caption{
{\bf Parameter values for simulations of the Izhikevich model.}}
\begin{tabular}{|l+c|c|c|c|}
\hline
 & $a$ & $b$ & $c$ & $d$ \\ \thickhline
  Figure~\ref{fig:bistability}B & 0.1 & 0.26 & $-60$ & 0\\ \hline
  Figure~\ref{fig:bistability}C & 0.1 & 0.2 & $-60$ & $-25$\\ \hline
  Figure~\ref{fig:spike-latency}B & 0.02 & 0 & $-65$ & 6\\ \hline
  Figure~\ref{fig:spike-latency}C & 0.02 & 0 & $-62$ & 0\\ \hline  
  Figure~\ref{fig:ADP}B & 0.5 & 0.6 & $-65$ & 6\\ \hline
  Figure~\ref{fig:ADP}C & 1 & 0.2 & $-60$ & $-15$\\ \hline  
\end{tabular}
\label{tab:Izhikevich}
\end{table}

The square-wave bursting in Figure~\ref{fig:MQIF-bursting}B was obtained using Eq~\eqref{eq:MQIF_us} with the parameters $C = 1$ ms, $\tau_s = 10$ ms, $\tau_{us} = 100$ ms, $V^0 = -40$, $V_s^0 = -38.4$, $V_{us}^0 = -50$, $\bar{g}_f = 1$, $\bar{g}_s = 0.5$, $\bar{g}_{us} = 0.015$, $V_{s,r} = -35$, $\Delta V_{us} = 3$, $I = 5$. The parabolic bursting in Figure~\ref{fig:MQIF-bursting}C was obtained using Eq~\eqref{eq:MQIF_uus} with the parameters $C = 1$ ms, $\tau_s = 10$ ms, $\tau_{us} = 100$ ms, $\tau_{uus} = 1000$ ms, $V^0 = -40$, $V_s^0 = -40$, $V_{us}^0 = -20$, $V_{uus}^0 = -50$, $\bar{g}_f = 1$, $\bar{g}_s = 0.5$, $\bar{g}_{us} = 0.1$, $\bar{g}_{uus} = 0.01$, $V_{s,r} = -25$, $\Delta V_{us} = 3$, $\Delta V_{uus} = 3$, $I = 110$.

The simulations in Figure~\ref{fig:MQIF-mod} use Eq~\eqref{eq:MQIF_us} with the same parameters as in Figure~\ref{fig:MQIF-bursting}B, except for $V_s^0$ and $V_{us}^0$, which were modulated. The values for $V_s^0$ were $-41$ (left), $-39$ (middle) and $-38.5$ (right). The values for $V_{us}^0$ were $-50$ (top) and $-54.5$ (bottom).

The f-I curves in Figure~\ref{fig:MQIF-type} were obtained using Eq~\eqref{eq:MQIF} with the same parameters as Figure~\ref{fig:MQIF-bursting}B, except for $V_s^0$, which was modulated. The values for $V_s^0$ were $-41$ (left), $-40$ (middle) and $-39$ (right).

\section*{Acknowledgements}
TVP received a fees scholarship from the Engineering and Physical Sciences Research Council (https://www.epsrc.ac.uk) under grant number 1611337. Both TVP and RS were supported by the European Research Council (https://erc.europa.eu) under the Advanced ERC Grant Agreement number 670645. The funders had no role in study design, data collection and analysis, decision to publish, or preparation of the manuscript.

\bibliography{../../references_bibtex}

\end{document}